# Low temperature phase diagram of hydrogen at pressures up to 380 GPa. A possible metallic phase at 360 GPa and 200 K

M. I. Eremets*, I.A. Troyan, A. P. Drozdov

*Max-Planck Institut fur Chemie, Chemistry and Physics at High Pressures Group*
*Postfach 3060, 55020 Mainz, Germany*
**e-mail: m.eremets@mpic.de*

**Two new phases of hydrogen have been discovered at room temperature in Ref.[1]: phase IV above 220 GPa and phase V above ~270 GPa. In the present work we have found a new phase VI at P≈360 GPa and T<200 K. This phase is likely metallic as follows from the featureless Raman spectra, a strong drop in resistance, and absence of a photoconductive response. We studied hydrogen at low temperatures with the aid of Raman, infrared absorption, and electrical measurements at pressures up to 380 GPa, and have built a new phase diagram of hydrogen.**

## Introduction

A strong motivation for study of solid hydrogen at high pressure is a predicted metallic state [2]. This metal is expected to have remarkable properties: high temperature superconductivity[3,4] or even a combined superconductivity and superfluidity[5] in case of a liquid ground state. However the predicted atomic state[2,6] of hydrogen still is not achieved. Some approximation of the metallic hydrogen was realized in $H_3S$ crystal under pressure where hydrogen sublattice is stabilized by sulfur atoms[7]. Pure hydrogen remains sustains in molecular state in phases I-III and in mixed atomic-molecular phase IV to the highest pressures of ~320 GPa[1,8-11].

Determination of the structures of the phases is difficult as X-ray scattering by hydrogen atoms is extremely weak. Only the hcp structure of the phase I is reliably established from single crystal X-ray diffraction studies[12]. Structure of other molecular phases II and III is not solved yet. Likely phase II structure of *P*21/*c* symmetry [13] as follows from combined neutron and X-ray diffraction studies.

The present knowledge on the structures of the phases of hydrogen is mostly from numerous calculations which take into account available optical data. It turned out that the *ab* *initio* predictions are very difficult because of the small energies of competitive phase, the search is limited by relatively small cells, and the real structure might be missing. There is a significant progress in the calculations (some recent refs: [14-17]). Hydrogen is very special difficult case for the calculations: atoms all corrections (zero point motion, anharmonicity and others) are large and should be included into consideration. DFT and other calculations have limitations. The diffusion quantum Monte Carlo (DMC) (QMC) method is generally regarded as the most accurate first-principles method appreciable for study of hydrogen. The recent QMC calculations [18] also include the temperature effects. However the QMC calculations still give very only qualitative agreement with the experimentally obtained phase diagram (Fig. 1).

In spite of the difficulties of determination of the structures from X-ray and neutron diffraction, the boundaries between the phases can be established accurately from Raman and infrared measurements. The low temperature phase diagram (phases I-III) was summarized in the review[19]. Next, two new phases (phases IV-V) were discovered at room temperature studies[1]. These phases were called as phases IV and IV' in the later publications[9,20,21]. The phases likely belong to the mixed molecular-atomic phases predicted by Pickard and Needs[22]. The domain of the new phases was extended to lower temperature[9,20] and high temperature range[23]. The high pressure phase IV and beyond were studied in numerous theoretical works[14,18,24-27].

The purpose of the present work is to build phase diagram of hydrogen in the low temperature domain (below room temperature) based on optical (Raman and IR spectroscopy) and electrical measurements at pressures up to 380 GPa. As a starting point for the study we used the phase diagram presented in Refs[19,28]. We assumed that the boundary between phases I and II is well established in numerous studies[29,30], and revisited the phase I-III boundary line extended it above room temperature, established domains of the phases IV and V. We present comprehensive study of the low temperature phase diagram at $P<380$ GPa.

At higher pressures we discovered the new, likely metallic low temperature phase (VI) at pressures $P>360$ GPa and $T<200$ K. The phase was detected by disappearance of the Raman modes, simultaneous with the drop of resistance, and the vanishing of a photoconductive response. We present data of one pressure run and are working on further verification of the new phase. In particular we are improving techniques for electrical measurements for proper measurements of the metallic and possibly superconducting state of the new phase. The major part of the results was reported at the Gordon conference on high pressure in 2014.

Finally we comment on the publication[31] appeared a week ago. A new phase at P>325 GPa at room and higher temperatures is claimed. Our data do not support this statement; the comments are at the end of text).

**Room temperature studies**

Here we extended study of phase transformations in hydrogen at room temperature obtained in the very first work[1] where two new phases were discovered. The first phase (phase IV) appears at P>220 GPa with the new Raman spectrum, and the next phase (phase V) at P~280 GPa (Ref[1] SI Fig. S7). The last phase was identified from changes in Raman spectra and drop of resistance. We present here more detailed and extended study (Fig. 1). At room temperature hydrogen is in phase I from the lowest pressures until at 202 GPa where it transforms to phase III as clearly is observed in the infrared measurements[32] (Fig. 2). Shortly, at 220 GPa, hydrogen transforms to phase IV with very characteristic Raman spectrum in the low frequency range <1300 $cm^{-1}$ (Fig. 3). At pressures above 270-280 GPa the low frequency Raman spectra of phase IV noticeably change (first documented in Ref.[1]): a new peak around 1100 $cm^{-1}$ arises (Fig. 3a-d). Other peaks gradually vanish except of the 300 $cm^{-1}$ peak which strongly decreases in intensity (Fig. 3b). We ascribe all these changes in the Raman spectra by entering phase V which coexists with phase IV in the 280-340 GPa pressure range where phonon branches at 700 $cm^{-1}$ and 1200 $cm^{-1}$ characteristic to phase IV remain. At higher pressures only two peaks of phase V around 1000 $cm^{-1}$ and 300 $cm^{-1}$ exist. Apparently phases IV and V do not differ significantly as the 300 $cm^{-1}$ mode remains the same, also the intensity and the frequency pressure dependence of hydrogen vibron does not change noticeably at the transition IV-V transition (except of maybe a subtle change of the slope of $\nu(P)$).

We comment here steps on this pressure dependence observed in numbers of runs (Fig. 3d)(see also Refs[20,31]). It is tempting to connect them with the pressure transitions in the 270 GPa and 320 GPa range. On the other hand the steps were not always observed and reproduced (see green line at Fig. 3d for comparison). More natural explanation seems to be that the steps are precursors of forthcoming failure of diamonds – diamonds break soon after the appearance of the steps. The pressure is determined from the diamond scale introduced in Ref.[33] (Fig. 10). In this work it was pointed out that the measured value is not real pressure inside the sample, but a reflection of the stresses in diamond adjoining to the sample . The stresses are complicated and depend on many parameters (geometry, gasket, cracks etc). Therefore, if the stresses redistribute before the failure so that the measured pressure seemingly increases, the pressure in the sample not necessarily increases. Anyway, a step on the pressure dependence of vibron frequency cannot be a reliable indication of a phase transformation in contrast to the claim in Ref[31] where the step on the $\nu(P)$ is one of the major evidences the transition at 325 GPa.

**Low temperature studies**

We established low temperature phase diagram by finding boundary between phases with the aid of Raman spectroscopy and electrical resistance measurements.

**I-III phase boundary**

The boundary between phases I and III has been determined from Raman and IR studies[19,28] but only up to ~230 K with large error bars at the highest temperatures. Therefore we revised data at low temperatures and extended the line to room temperatures (Fig.1). We used Raman spectroscopy for detection of a transition between different phases. Typically we recorded Raman spectra at a constant pressure but varying temperature. At the low pressure range (158 GPa, Fig. 4) transition between phases I and III is easily detected. The temperature dependence of the energy of the vibron is very different in two phases, the frequency of the hydrogen vibron drops at ~40 $cm^{-1}$ at the transition, the width of the vibron significantly increases at transition to phase III (insert in Fig. 4a). The transition occurs in the temperature range of ~30 K. At higher pressures (see, for instance 185 GPa, Fig. 4b) the gap in the frequencies of the phases is less evident (insert in Fig. 4b) but the point of the transformation is obviously determined as the kink at the temperature dependences of the frequency of the vibron. At this point the width of the vibron increases at transition to phase III (upper insert in Fig. 4b).

**III-IV phase boundary**

Phase IV has been discovered in Ref [1] in the pressure run at room temperature: above 220 GPa new Raman peaks appeared manifesting a new phase. Here we performed temperature measurements and established the phase boundary between phases III and IV (see also Refs[9,11]).

In the first experiment we loaded the hydrogen sample at room temperature at 160 GPa in the phase I and cool it down to 180 K. We observed the known transition to phase III (Fig. 4)[19]. At this temperature we intended to increase pressure above 280 GPa to search for metallic state but were able to increase pressure not more than 242 GPa and were forced to warm up the sample. To our surprise, a sharp transition occurred only approaching the ambient temperature in the ~268-273 K range. We found that the Raman spectrum of the new phase is typical for phase IV [1]. After this observation we systematically studied the III-IV transition and built the phase line. Typically we increased pressure at room temperature to enter phase IV and then cooled the sample while keeping pressure constant and observed changes in the Raman spectra. At low pressures of 217 GPa (Fig. 5) this transition is apparent in spite of only a weak peak at ~350 $cm^{-1}$ belonging to phase IV appears but the transition is clearly identified from the big jump of the vibron frequency at ~60 $cm^{-1}$ in the temperature range of ~4 K (Fig. 5b). At higher pressures the jump in the vibron frequency strongly increases (Fig. 5d), reaching > 300 $cm^{-1}$ at 252 GPa. This is one order of magnitude more than the jump of vibron frequency by crossing to I-III line ($\Delta\nu$ ~30 $cm^{-1}$). This big difference indicates a significant difference of the structure of phases IV and III in contrast to phases I and III (both are close to the *hcp*).

It was instructive to cross the III-IV line at increasing pressure at constant temperature (250 K). Starting from 230 GPa at room temperature we cooled down the sample, observed transition to phase III at ~270 K, then cooled down to 250 K, then increased pressure at this temperature until crossed again the III-IV boundary (Fig. 5e) and entered phase IV. At the LT loading vibron followed the well-known pressure dependence [8,34] – red points at Fig. 5e. Above 250 GPa after the crossing III-IV line frequency of vibron jumped to the pressure dependence of vibron in phase IV[1]. This experiment makes a bridge between LT[8] and surprisingly different RT Raman measurements [1]. It is clear from the insert in Fig. 5e that the jump in the frequency of the vibron strongly increases with pressure.

**Triple point between phases I, III, and IV**

I-III and III-IV phase boundaries intersects at 208 GPa and T=308 K. To check, if this is a new triple point we searched for a new phase boundary at temperatures above this point. For that we applied pressure of 180 GPa and then increased temperature from 298 K to 324 K at this pressure. The pressure did not change during the heating according to the diamond edge scale[33]. We crossed the phase diagram by changing pressure from 170 to 220 GPa at a constant temperature of 324 K (Fig. 6). A phase transformation was observed at 208 GPa as it is evident from the kink on the pressure dependence of the vibron and a dramatic increase of width of the vibron above this pressure (Fig. 6).

**III-V phase line**

Transitions at cooling from phase V to phase III (Fig. 7) are similar to IV-III transformations (Fig. 5). The Raman spectra in phase III are richer at P>~270 GPa. Likely the sharpening of the spectra is connected with lowering of temperature of the phase transformation, and the phase III remains. However a subtle orientation transition is not excluded. There is an important change of the phase line: it becomes nearly horizontal at the highest pressures of ~350 GPa (Fig. 1).

**Transition to the new phase VI**

At higher pressures of 360-370 GPa we observed (Fig. 8a) that the Raman peaks from the phase V disappeared at cooling below 220 K. This result was reproduced in another cooling run (Fig. 8b) with smaller temperature steps and the temperature of disappearance of the Raman spectra was localized at 204 K. Moreover at the same time and the same temperature we observed a strong drop in the resistance (Fig. 9a). The nearly temperature independent resistance and vanishing of photoconductive response suggests that the new phase could be metallic. The measured resistance (about MOhm) however is too high for metal. This high value probably is a matter of the new, not well established ac measurements and should be checked in future experiments. An apparent drawback of the measurements is quasi-four probe scheme (Fig.11). Thus a contact resistance between the sample and electrodes is included into measurements. The true four-probe arrangement as has been done in Ref.[35] is required. Apparently all these improvements will need a lot of efforts and time keeping in mind the very high pressure range. Therefore we present the electrical experiment in the present preliminary form.

We should mention that electrical measurements in the present study differ from used in previous work[1]: instead of DC measurements we used an impedance spectrometer to measure conductivity at different frequencies in the range of 10-100 000 Hz. Likely because of this we did not observe drops in resistance observed in Ref. [1] in the range of 280 GPa – transition into phase V - probably because of different mechanism of conductivity (protonic or electronic). We did not compare directly the DC and the AC methods yet.

Thus at pressures of ~350 GPa we observed a transition to a new low temperature conductive, likely metallic phase as follows from the sharp disappearance of the Raman spectra, the drop of resistance, and the absence of a photoconductive response.

Finally we comment the claimed discovery of the new phase at P>325 GPa at room and higher temperatures in Ref. [31]. This publication appeared a week ago and motivated us to finalize our nearly finished present article. Our data do not support the statement of the new phase claimed in Ref.[31] (see also comments in the "room temperature studies" in the present article). The main arguments in Ref.[31] are the following:

(1) appearance of a step at the pressure dependence of the vibron (observed in fact only in one run). However the strep cannot be good evidence: we observed such steps at different pressures at a number of runs; but the steps are not well reproduced, in some runs there were no steps (Fig. 3e). Likely the steps reflect changes of the stresses in the diamond tip prior the forthcoming failure of diamonds which happened soon after the appearance of the step.

(2) increase of the full-width at half-maximum for the 300 $cm^{-1}$ mode. We do not support a significant increase of the full-width at half-maximum: for instance it is 74 $cm^{-1}$ at 310 GPa and 86 $cm^{-1}$ at 375 GPa in our measurements. This slight increase in fact is not clear because the accuracy is poor. In particular extraction of a noticeable background is required, and this is uncertain procedure with a possible systematic error due to background increasing with pressure.

(3) disappearance of some low frequency modes. This is also a poor indication of phase transition as the intensity of these modes decrease gradually starting from ~280GPa (Phase V).

In our opinion, instead of introducing the new phase in Ref[31] it is more naturally to suppose that two phase IV and V (in our definition, or phases IV and IV' in definition in Ref[31]) coexist in the 270-320 GPa range and phase V remains at higher pressures. The coexistence of phases is common for hydrogen (Figs 4.5,7).

Acknowledgements. Support provided by the European Research Council under the 2010 Advanced Grant 267777 is acknowledged. The authors appreciate participation of P. Naumov in the early stage of this work.

# Figures

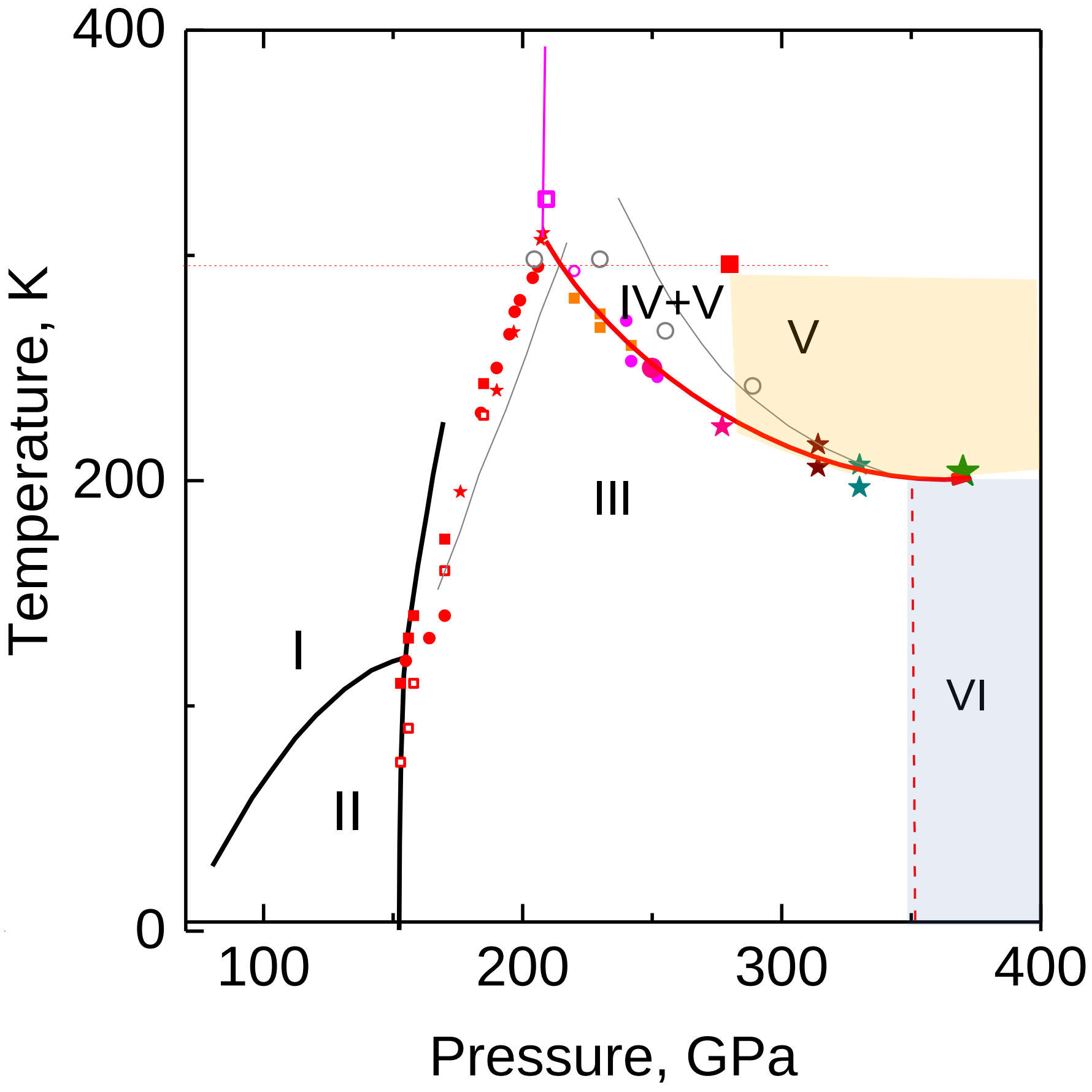

Fig. 1. The phase diagram of hydrogen. Letters I-VI indicate domains of different phases. Phase boundaries between phases I-III (black lines) are taken from Ref. 16. Red horizontal line is the room temperature run where red square indicate a phase transition (Ref. 1). Color points (red orange, and magenta) – present work - indicate points of phase transformations derived from changes in Raman and infrared spectra. Linear interpolation of these data points gives the triple point is at 208 GPa and 305 K. Emergence of phase III in the room temperature run (dot line) is apparent in infrared absorption measurements (Fig. 2). Grey circles are data from Ref. 17. 18. Note, that in both sets of data the measured pressures are overestimated - shifted to higher pressures relative to our results. Domains of phases V and VI are tentatively indicated by yellow and grey colors. In the 220-270 GPa pressure range likely there is a mixture of phases IV and V.

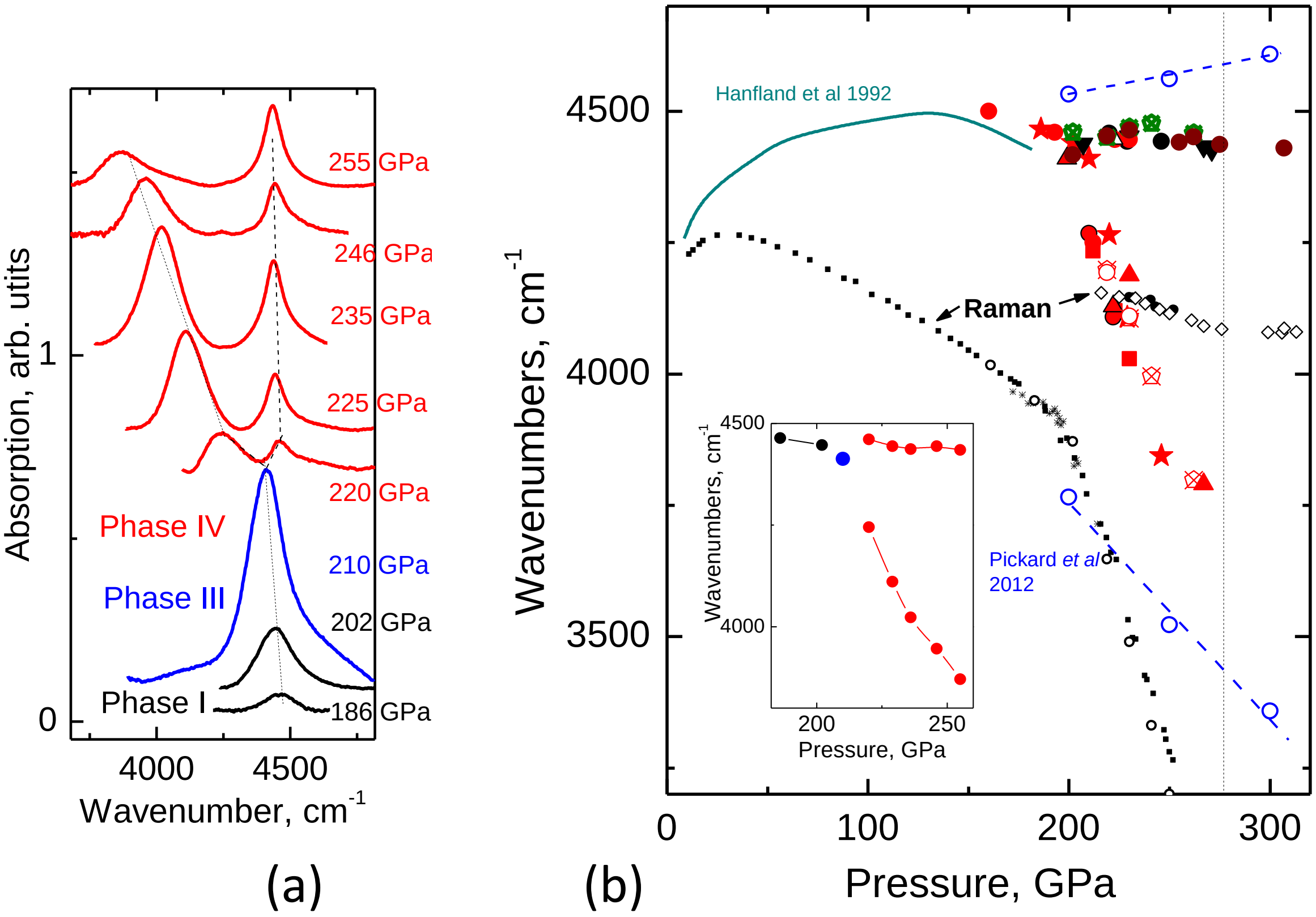

Fig. 2. The pressure dependence of IR and Raman vibrons at room temperature. (a) The pressure dependence of IR vibron spectra. (b) The pressure dependence of frequency of Raman and infrared vibrons. Insert shows transformations above 220 GPa when hydrogen transforms from phase I (black points) to phase III (blue points), and then to phase IV. See Ref. 31 for further details.

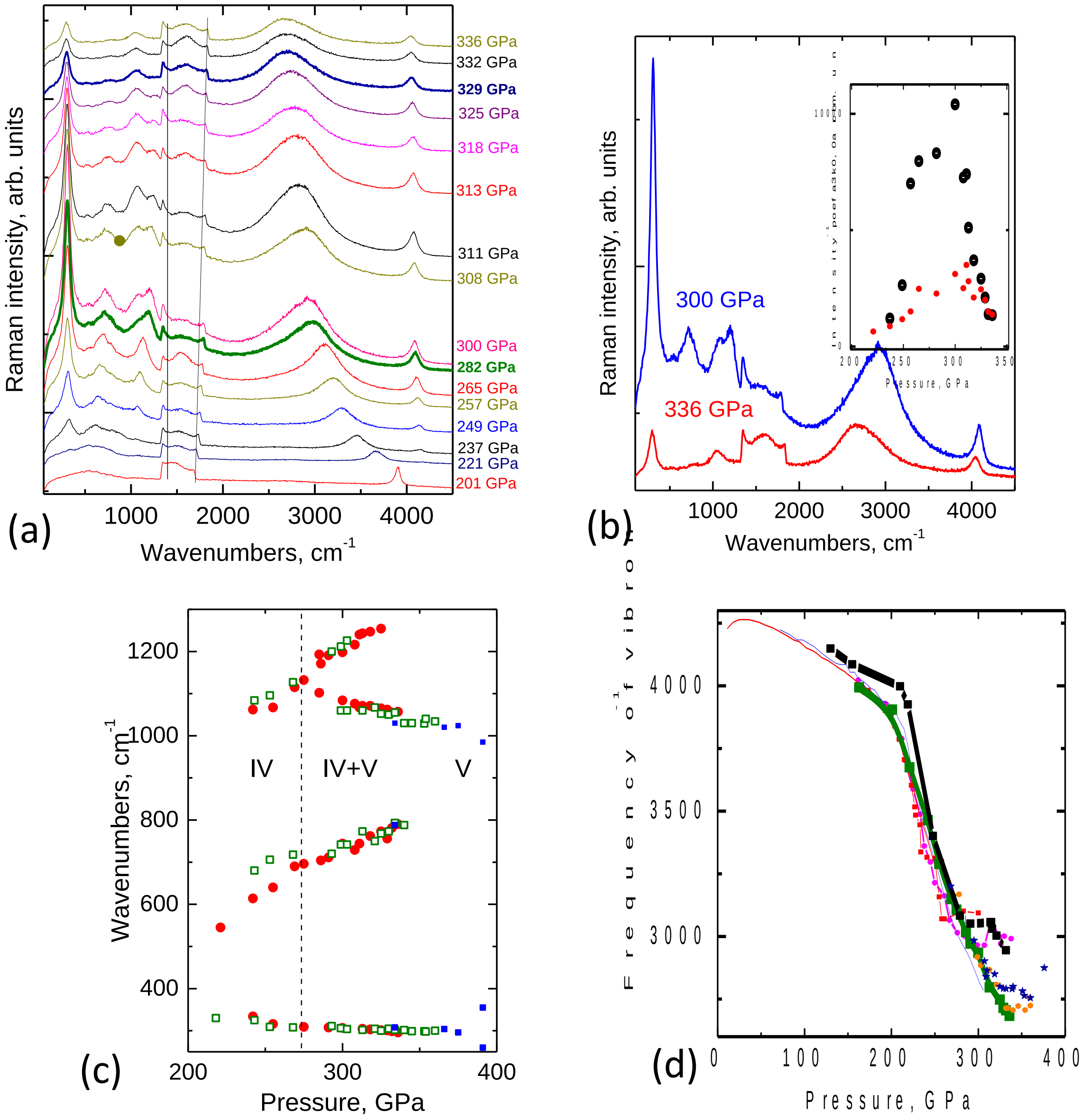

Fig. 3. Raman spectra of hydrogen at pressures at room temperature. (a) Evolution of Raman spectra with pressure. Black lines indicate the region of Raman signal from the stressed tip of diamond anvil. The high frequency edge shifts with loading and its position was used to determine pressure according to Ref. 33. Transformation from phase IV to new phase (V) happens at pressures ~270 GPa. Above this pressure the low frequency spectrum changes: the 1100 $cm^{-1}$ splits. The 300 $cm^{-1}$ peak strongly decreases in intensity as obvious in comparison of two spectra (b) and the insert where intensity of the 300 $cm^{-1}$ is plotted by black circles, vibron decreases in intensity not as dramatically at higher pressures (red circles). (c) The pressure dependence of the low frequency modes in different runs indicated by different colors. The phase V enters at P>270 GPa, and the phases IV and V likely coexist in the 270-320 GPa range. At higher pressures only phase V is observed in the Raman spectra with two characteristic modes at ~300 $cm^{-1}$ and ~1000 $cm^{-1}$.

(d) Pressure dependence of vibron for different runs. At some runs steps on the pressure dependence ν(P) are formed at pressures about 270-280 GPa and above ~300 GPa. Probably they reflect phase transformations in the sample which can be explained, if volume of the sample drops. In this case pressure in the sample stays until the transition will finish while the load (the measured stresses in the diamond tip) increases. On the other hand these steps are not well reproduced. For instance, olive line shows no steps to the highest pressures. Likely the steps reflect changes of the stresses in diamond anvils prior the failure of the anvils.

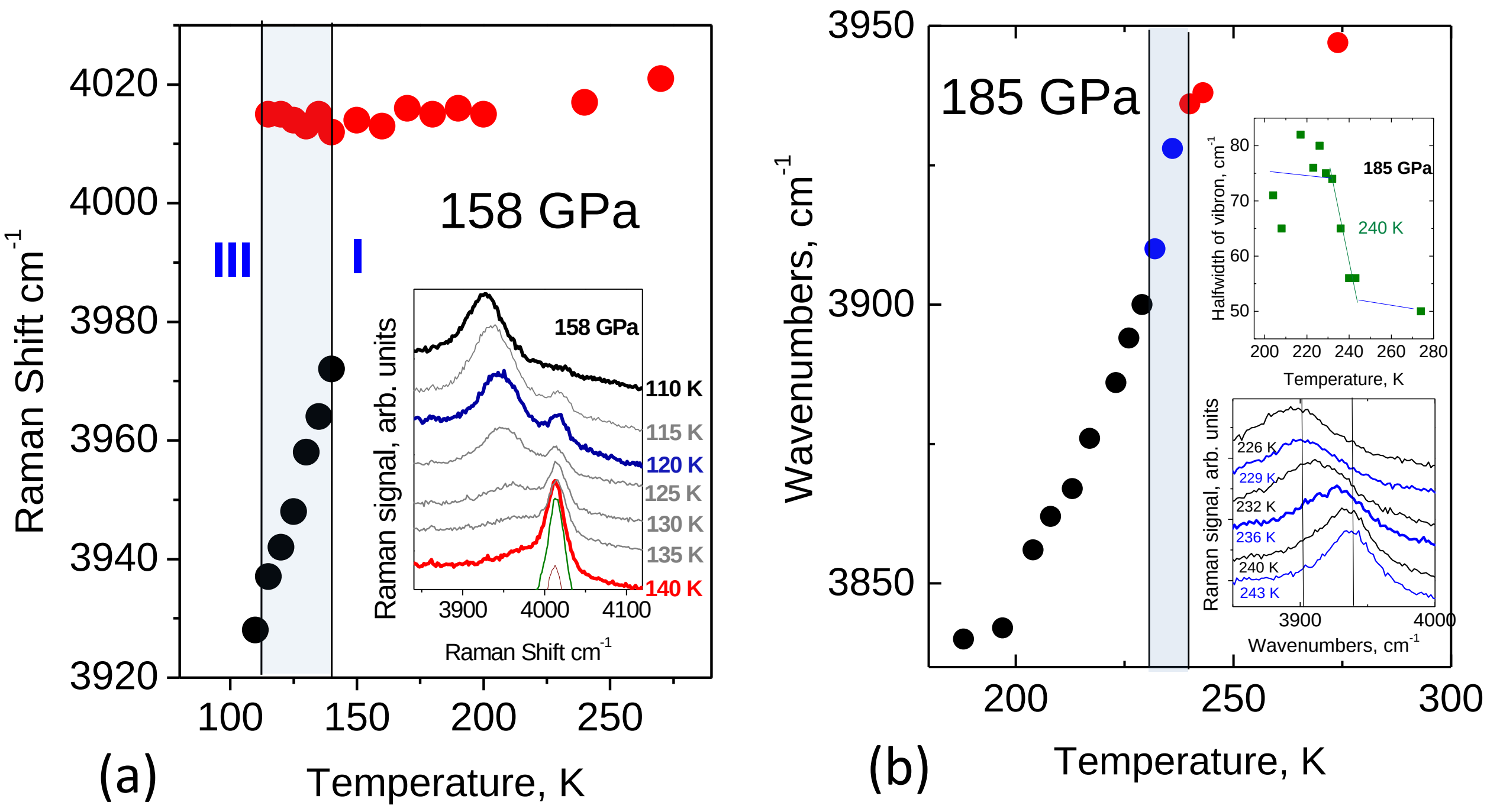

Fig. 4. Phase transformation from phase I to phase III at different pressures as measured from Raman spectra of the vibron of molecular hydrogen. (a) At cooling from room temperature at 158 GPa position of the vibron nearly does not change until 140 K when the vibron of the phase III appears at lower frequencies. In phase III, the vibron frequency strongly depends on temperature, the half width of vibron is significantly higher as can be seen from the spectra in insert. In the 110-140 K temperature range (dashed domain) the I and III phases coexist. (b) At 185 GPa, the I–III phase transition is obvious from the kink at 240 K and the strong change of the temperature dependence of vibron frequency. The jump in frequency during this transition as well as the increase of the half width (shown in insert) is significantly smaller in comparison with the those at 158 GPa and further reduce at higher pressures and temperatures.

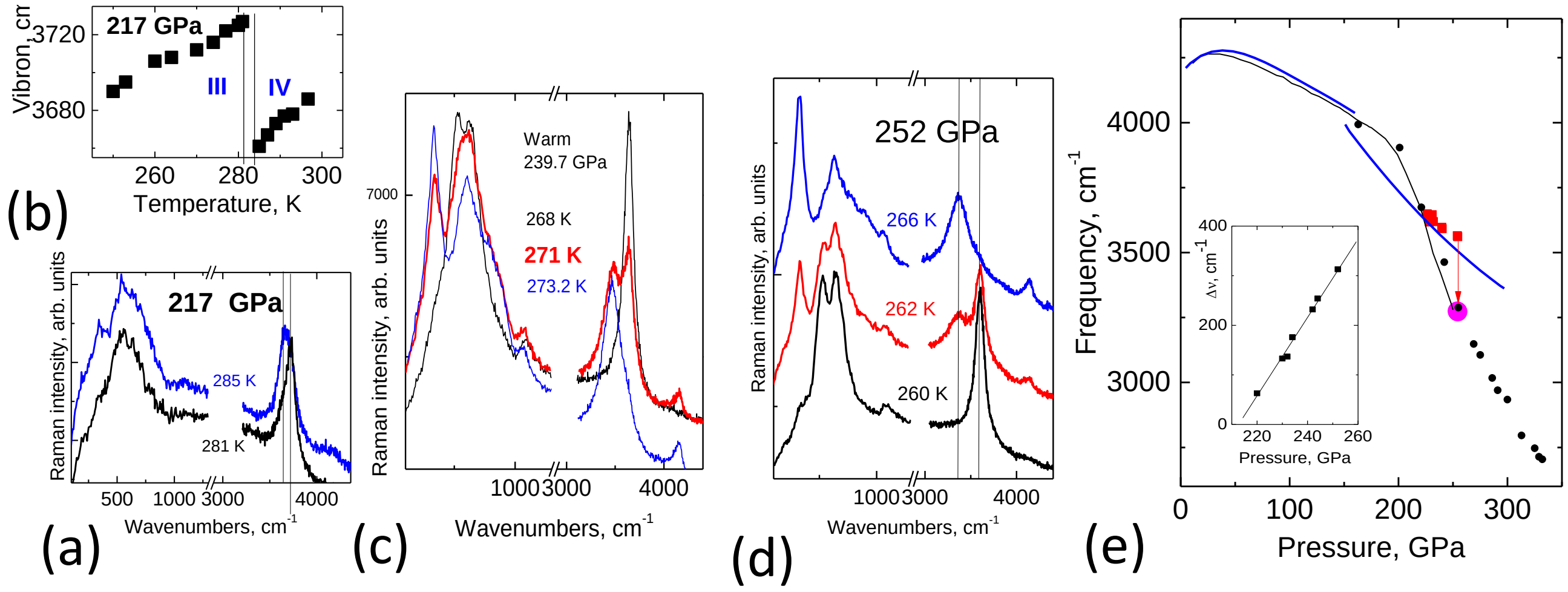

Fig. 5. Changes of Raman spectra of hydrogen at transition between phases III and IV. The III-IV transition develops in Raman spectra at cooling at 217 GPa: a small 300 $cm^{-1}$ peak appears, and frequency of vibron jumps as shown in detail in (b). (c-d) At cooling at higher pressures of 240 GPa and 252 GPa this transition is pronounced both in the range of phonon and vibron excitations. In the temperature range of 6 K phase IV transforms to phase III: characteristic low frequency peaks of phase IV vanish, as well as the second vibron peak at 4200 $cm^{-1}$; vibron frequency strongly decreases. (e) Pressure dependence of the frequency of hydrogen vibron in the phase III (blue line) and phase IV (black line and points). The red arrow indicates the jump of vibron at the transition from phase III to IV at 252 GPa and 262 K. The jump in the vibron frequency strongly increases with pressure (insert).

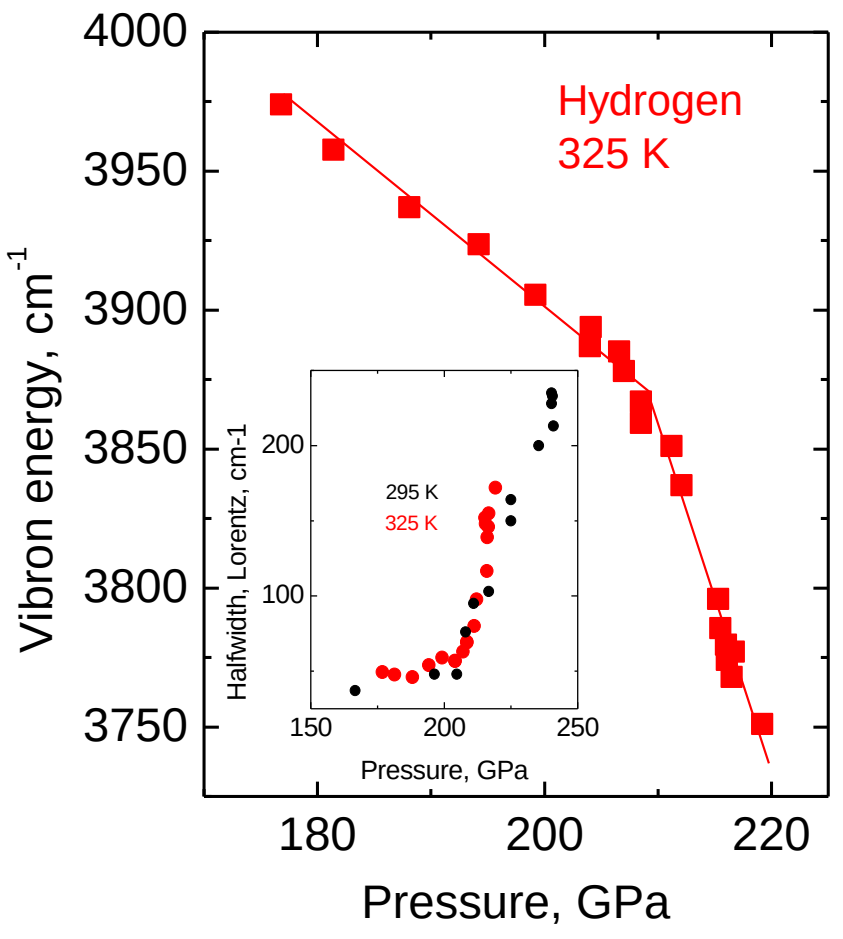

Fig. 6. Transition between phases I and IV measured from the pressure dependence of vibron at 325 K. In this run pressure was increased, while the temperature of 325 K maintained constant. Pressure was determined with the diamond edge scale Ref. 33. At pressure above 209 GPa the derivative of pressure dependence of vibron changes, and the width of vibron dramatically increases.

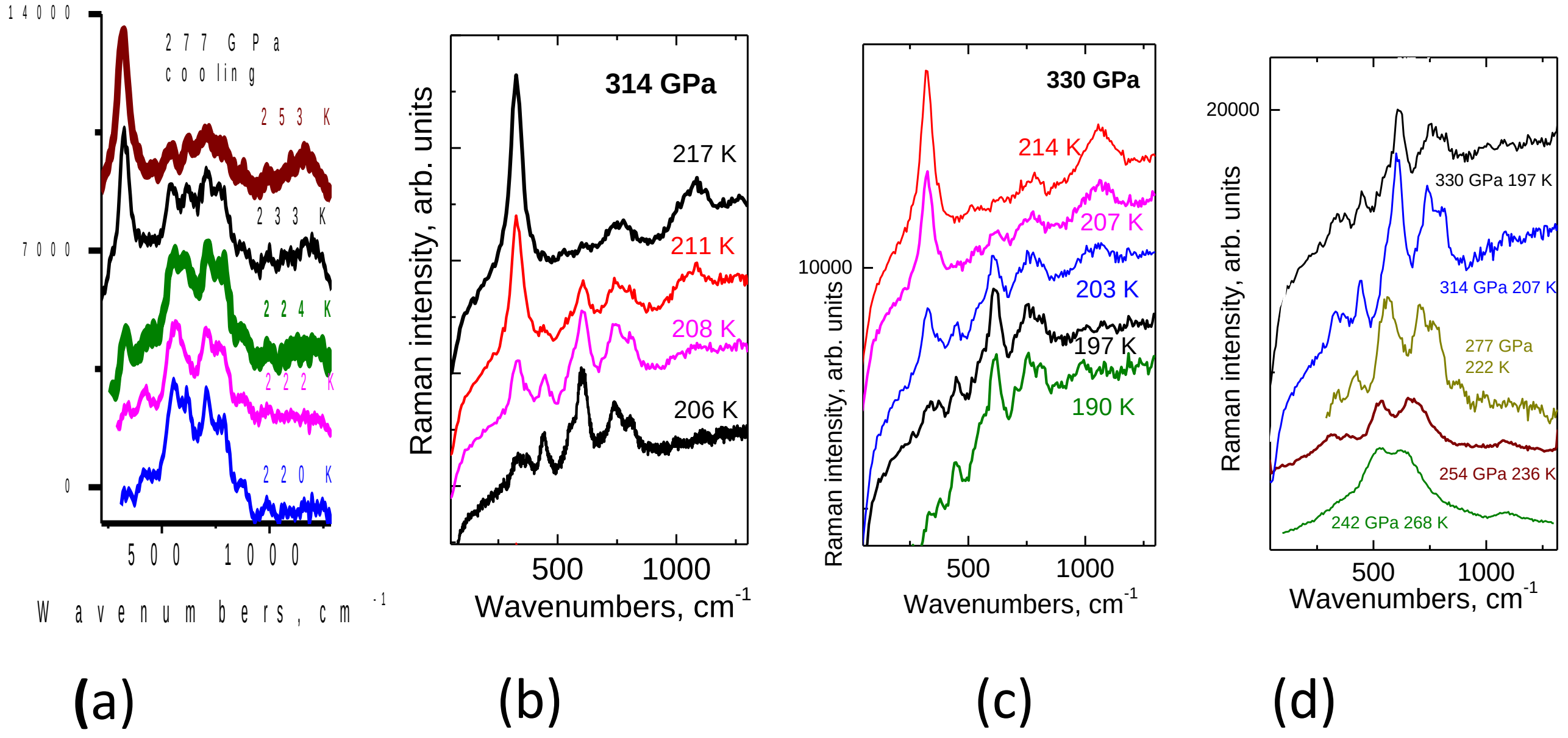

Fig. 7. Raman spectra at transition between phases III and V. (a-c) phase transformations at 277 GPa, 314 GPa and 330 GPa during cooling. The transitions happened in small temperature range of less than 10 K. (d) comparison the Raman spectra at different pressures. Likely the spectra relate to the same phase III. They become however more pronounced with pressure and lowering of temperature.

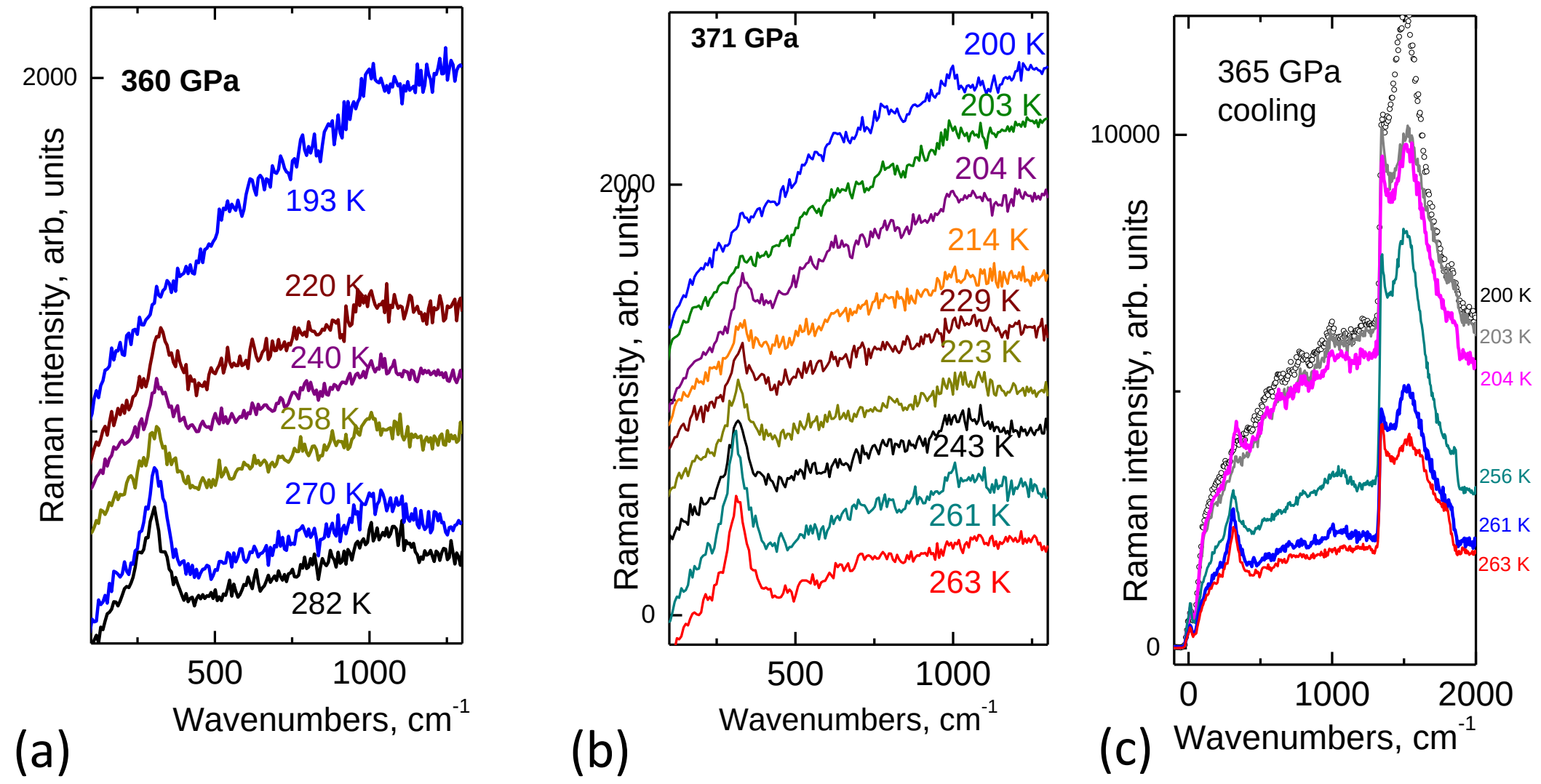

Fig. 8. Raman spectra at cooling of hydrogen at the highest pressures. (a) At the first cooling (at 360 GPa) the spectra characteristic for phase V transform to nearly featureless spectrum below 220 K. The spectra are shifted vertically for more clear presentation of changes in the spectra as well in (b). (b) At the second cooling (pressure increased to 370 GPa in between two cooling) the transition to the featureless spectrum was reproduced and the temperature of the transition was localized at 203 K. The resistance was measured (Fig. 9a) simultaneously with the measurements of Raman spectra. Only Raman spectra were measured in (b) as the electrical leads were broken after the first cooling. (c) The same spectra as (b) but without vertical shift to show that in spite of the increasing luminescence the 300 cm$^{-1}$ K is clearly seen at 204 K but disappears at 203 K and below.

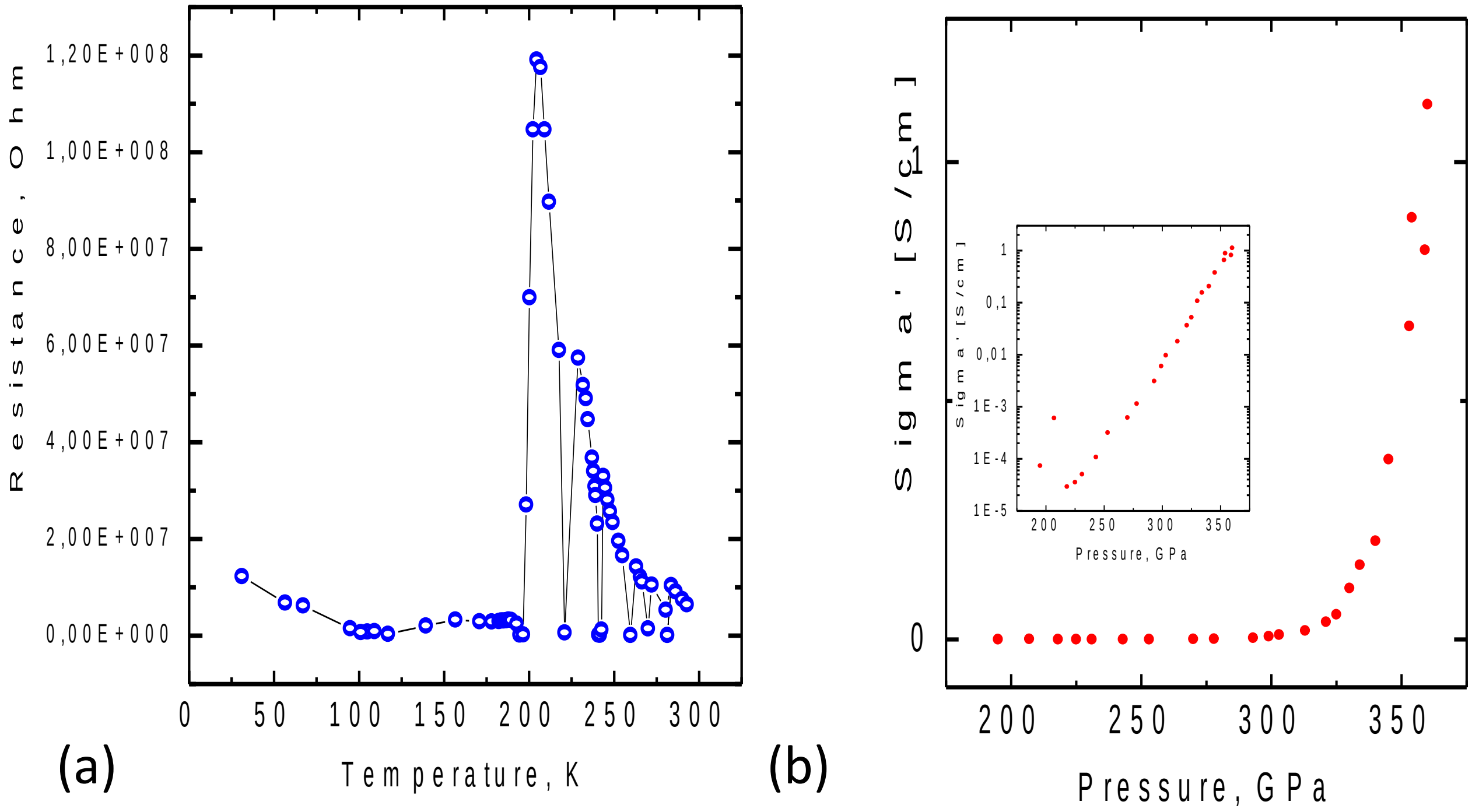

Fig. 9. Electrical measurements. (a) At cooling at 360 GPa the resistance first increased showing semiconducting behavior (drops in the resistance are from illumination by laser during the Raman measurements). At 200 K the resistance strongly dropped, at the same temperature Raman spectra dramatically changed too and characteristic peaks disappeared (Fig. 8). At lower temperatures the resistance is nearly temperature independent. The absolute value of resistance is too high for metal however. Probably the high resistance belongs not to the sample but can be explained by a contact sample-electrode resistance included in the measurements.

(b) The pressure dependence of conductivity measured at room temperature shown in normal and logarithmic (insert) scale. The conductivity appears at P> 220 GPa as soon as hydrogen transforms to phase IV.

# Methods

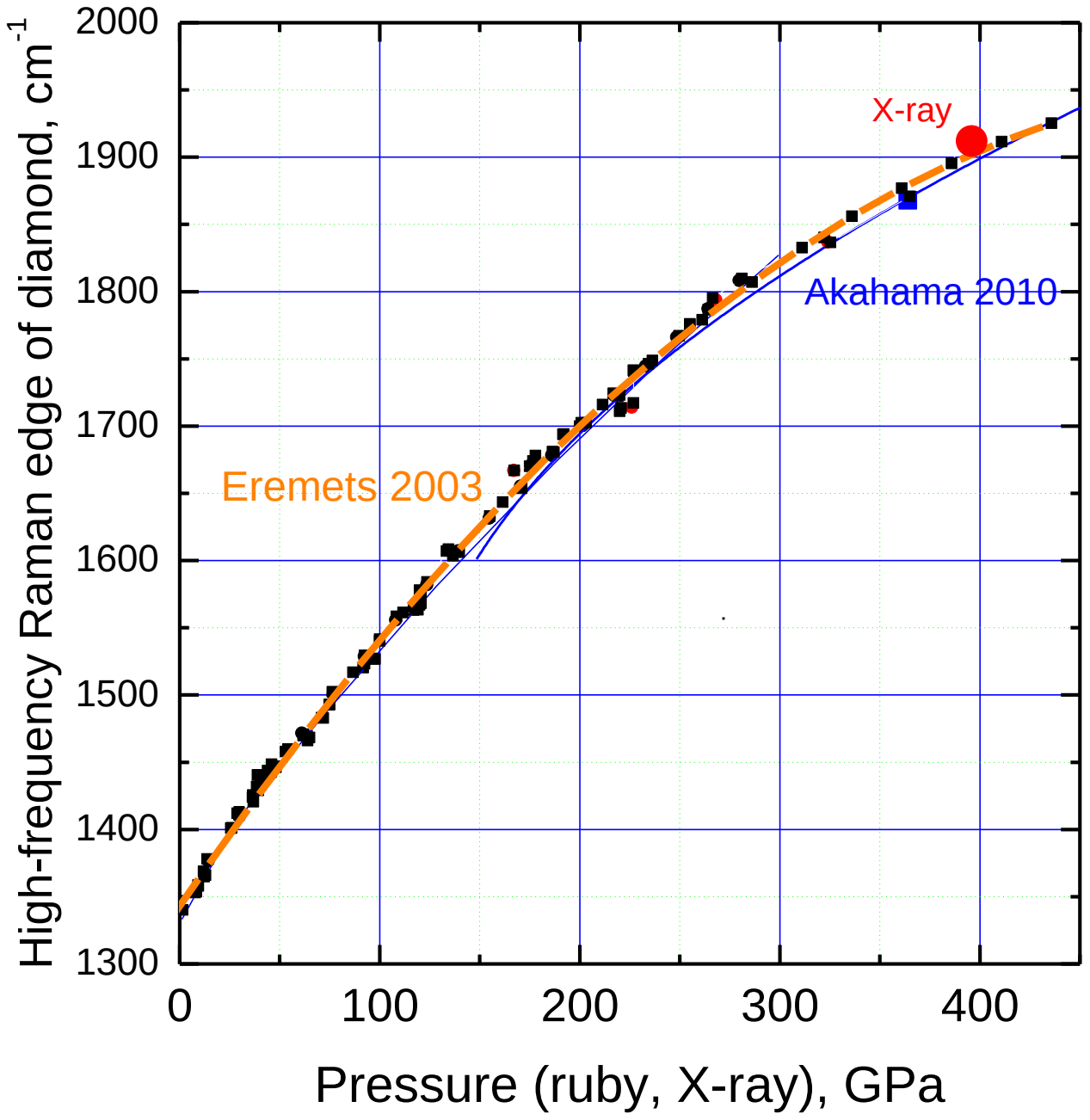

Fig. 10. The pressure calibration: the pressure measured by ruby scale and X-ray diffraction standards (gold) are plotted against the Raman signal from the stressed diamond anvils measured at the middle of the high frequency edge- orange line. It is based on the Ref. 33 data which were further extended to higher pressures. The highest measured point (red) is 396 GPa (X-ray from gold measured at DESY) which corresponds to 1912 $cm^{-1}$ of the diamond edge measured at the same time. Calibration by Akahama et al (Ref. 35) is shown by blue line for comparison, It gives slightly high values of pressure 5- 10 GPa in the 300-400 GPa range.

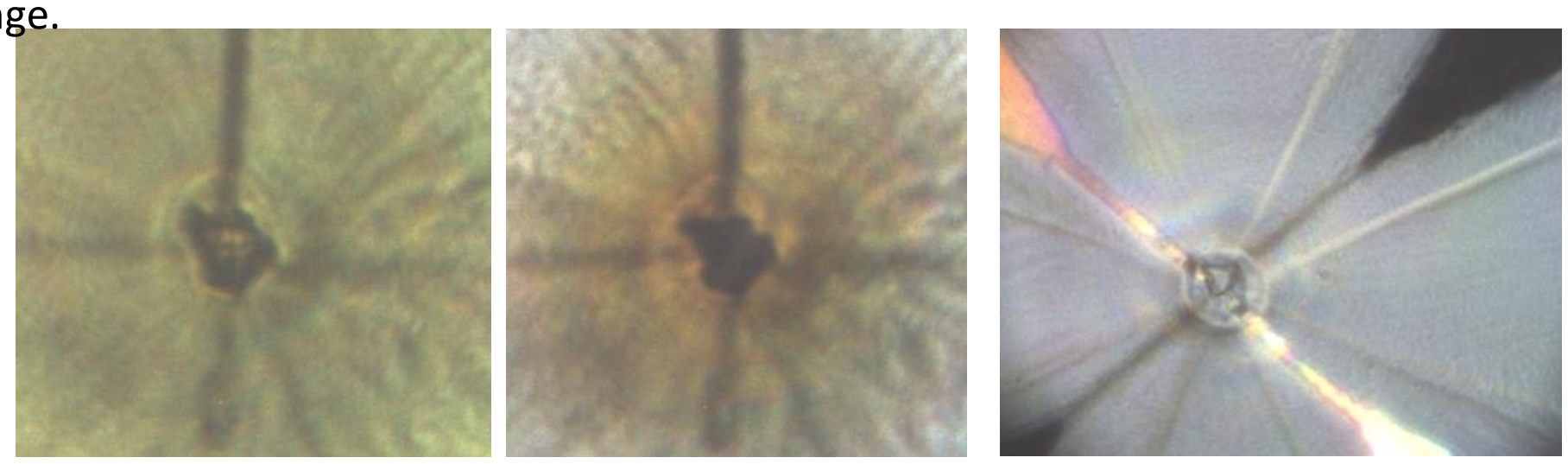

(a) 227 GPa (b) 318 GPa (c) 260 GPa

Fig. 11. Electrical measurements. The techniques is similar to Ref. 1. Two crossed electrodes are seen at the photographs. One electrode is sputtered at one anvil, second – on the opposite anvil. A measured electrical current appeared at pressure above ~220 GPa and increased with pressure. Another sample is shown in (c). Typically culet size was about 20 μm.

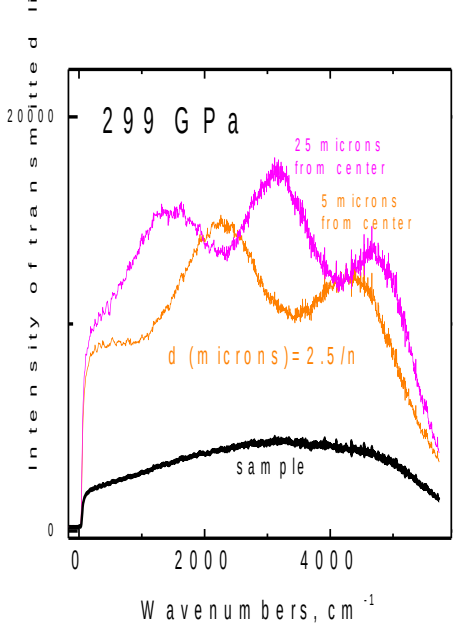

Fig. 12. The estimated thickness of the sample of hydrogen at 299 GPa from measurements of light transmitted two anvils. The culet forms a Pabry-Perrot resonator which produced interference fringes in the transmission spectra. The sample itself is opaque at 299 GPa but surrounding gasket (epoxy/MgO) is transparent. The thickness of the gasket increases as the distance from the sample increases. The measurements at the point close to the sample gives an estimation of its thickness as ~2 μm taking into account that the refractive index of the gasket may be about 1.5.